\documentclass[useAMS,usenatbib]{mn2e}
\usepackage{graphicx,fleqn,fix2col}
\DeclareGraphicsExtensions{.eps,.ps,.eps.gz,.ps.gz,.eps.Z}
\title[ULTRACAM observations of two accreting white dwarf pulsators]{ULTRACAM observations of two accreting white dwarf pulsators}

\author[C.M.~Copperwheat et al.]{C.M.~Copperwheat$^{1}$, T.R.~Marsh$^{1}$, V.S.~Dhillon$^{2}$, S.P.~Littlefair$^{2}$, \newauthor P.A.~Woudt$^{3}$, B.~Warner$^{3}$, D.~Steeghs$^{1}$, B.T.~G{\"{a}}nsicke$^{1}$ and J. Southworth$^{1}$\\
$^{1}$ Department of Physics, University of Warwick, Coventry, CV4 7AL, UK\\
$^{2}$ Department of Physics and Astronomy, University of Sheffield, S3 7RH, UK\\
$^{3}$ Department of Astronomy, University of Cape Town, Rondebosch 7701, South Africa
}

\date{Received: }

\begin{document}

\newcommand{\dg} {^{\circ}}
\outer\def\gtae {$\buildrel {\lower3pt\hbox{$>$}} \over
{\lower2pt\hbox{$\sim$}} $}
\outer\def\ltae {$\buildrel {\lower3pt\hbox{$<$}} \over
{\lower2pt\hbox{$\sim$}} $}
\newcommand{\ergscm} {erg s$^{-1}$ cm$^{-2}$}
\newcommand{\ergss} {erg s$^{-1}$}
\newcommand{\ergsd} {erg s$^{-1}$ $d^{2}_{100}$}
\newcommand{\pcmsq} {cm$^{-2}$}
\newcommand{\ros} {{\it ROSAT}}
\newcommand{\xmm} {\mbox{{\it XMM-Newton}}}
\newcommand{\exo} {{\it EXOSAT}}
\newcommand{\sax} {{\it BeppoSAX}}
\newcommand{\chandra} {{\it Chandra}}
\newcommand{\hst} {{\it HST}}
\newcommand{\subaru} {{\it Subaru}}
\def\rchi{{${\chi}_{\nu}^{2}$}}
\newcommand{\Msun} {$M_{\odot}$}
\newcommand{\Mwd} {$M_{wd}$}
\newcommand{\Mbh} {$M_{\bullet}$}
\newcommand{\Lsun} {$L_{\odot}$}
\newcommand{\Rsun} {$R_{\odot}$}
\newcommand{\Zsun} {$Z_{\odot}$}
\def\Mdot{\hbox{$\dot M$}}
\def\mdot{\hbox{$\dot m$}}
\def\mincir{\raise -2.truept\hbox{\rlap{\hbox{$\sim$}}\raise5.truept
\hbox{$<$}\ }}
\def\magcir{\raise -4.truept\hbox{\rlap{\hbox{$\sim$}}\raise5.truept
\hbox{$>$}\ }}
\newcommand{\mnras} {MNRAS}
\newcommand{\aap} {A\&A}
\newcommand{\apj} {ApJ}
\newcommand{\apjl} {ApJL}
\newcommand{\apjs} {ApJS}
\newcommand{\aj} {AJ}
\newcommand{\pasp} {PASP}
\newcommand{\apss} {Ap\&SS}
\maketitle

\begin{abstract} 
 In this paper we present high time-resolution observations of GW Librae and SDSS J161033.64-010223.3 (hereafter referred to as SDSS 1610) -- two cataclysmic variables which have shown periodic variations attributed to non-radial pulsations of the white dwarf. We observed both these systems in their quiescent states with ULTRACAM on the VLT and the University of Cape Town Photometer on the SAAO 1.9m telescope, and detect the strong pulsations modes reported by previous authors. The identification of further periodicities in GW Lib is limited by the accretion-driven flickering of the source, but in the case of SDSS 1610 we identify 
several additional low-amplitude periodicities. In both sources we find the pulsation modes to be stronger in amplitude at bluer wavelengths. In the case of SDSS 1610, there is evidence to suggest that the two primary signals have a different colour dependence, suggesting that they may be different spherical harmonic modes. We additionally observed GW Lib during several epochs following its 2007 dwarf nova outburst, using ULTRACAM on the VLT and the Auxiliary Port Imager on the William Herschel Telescope. This is the first time a dwarf nova containing a pulsating white dwarf has been observed in such a state. We do not observe any periodicities, suggesting that the heating of the white dwarf had either switched-off the pulsations entirely, or reduced their relative amplitude in flux to the point where they are undetectable. Further observations eleven months after the outburst taken with RATCam on the Liverpool Telescope still do not show the pulsation modes previously observed, but do show the emergence of two new periodic signals, one with a frequency of $74.86 \pm 0.68$ cycles/day ($P = 1154$s) and a $g'$-band amplitude of $2.20$\% $\pm 0.18$, and the other with a frequency of $292.05 \pm 1.11$ cycles/day ($P=296$s) and a $g'$ amplitude of $1.25$\% $\pm 0.18$. In addition to the WD pulsations, our observations of GW Lib in quiescence show a larger-amplitude modulation in luminosity with a period of approximately $2.1$ hours. This has been previously observed, and its origin is unclear: it is unrelated to the orbital period. We find this modulation to vary over the course of our observations in phase and/or period. Our data support the conclusion that this is an accretion-related phenomenon which originates in the accretion disc.  
\end{abstract}

\begin{keywords}
stars: individual: GW Librae, SDSS J16103.64-010223.3 --- stars: dwarf novae --- stars: oscillations --- stars: white dwarfs
\end{keywords}

\section{INTRODUCTION}  

Cataclysmic variable stars (CVs: \citealt{Warner95}) provide examples of white dwarfs (WDs) accreting from low mass companions and are the progenitor class of classical novae. Dwarf novae (DNe) are a subset of CVs which feature periodic outbursts, thought to be the result of thermal instability in the accretion disc leading to accretion at rates far in excess of the rate in quiescence. The emission from CVs is generally dominated by these accretion processes, making it difficult to probe the WD itself. Some isolated WDs show periodic variations which have been attributed to non-radial pulsations of the WD at surface temperatures between $11,000$ and $13,000$K \citep{Gianninas06a}. These are termed DAV WDs, or ZZ~Ceti stars (see \citealt{Bradley98} for a review). These WDs are relatively cool and lie within a region in the $T_{eff}$ -– $\log g$ plane termed the ZZ~Ceti instability strip \citep{Gianninas06a}. 

In recent years photometric observations of some DNe during quiescence have revealed the accreting analogues of the ZZ~Ceti WDs. The first CV of this type was the $\sim$$17$ mag DN GW~Librae (GW~Lib) \citep{Warner98,vZyl00}. A spectroscopic period of $76.78$ min has been reported by \citet{Thorstensen02} for this source, which makes it one of the shortest orbital period CVs known. \citet{vZyl00,vZyl04} presented amplitude spectra of GW~Lib from observing campaigns conducted during 1997, 1998 and 2001. They found the dominant pulsation modes to be clustered at periods near 650, 370 and 230 seconds. Observations at UV wavelengths showed these same pulsation modes \citep{Szkody02}. Further examples of accreting pulsators have been discovered largely as a result of the Sloan Digital Sky Survey (SDSS; \citealt{Szkody07}). Of the CVs discovered by the SDSS survey, the first found to contain a pulsating WD was SDSS J161033.64-010223.3 (\citealt{Szkody02}, SDSS 1610 hereafter). \citet{Woudt04} reported high-speed photometry of this source, taken with the University of Cape Town CCD Photometer mounted on the 74-inch Radcliffe reflector. They measured an orbital period of $80.52$ min, and non-radial pulsations with principal modes near $606$ and $345$ seconds. Signals at $304$ and $221$ seconds were also discovered. These frequencies of these modes suggest they are respectively a harmonic of the first mode and a linear combination of the principal modes.

Stellar pulsations in CVs have huge potential as probes of the WDs. The outer layers of the WD are modified by accreting solar abundance material at $\sim$$10^{-10}$\Msun/yr and the ejection of this accreted material via nova eruptions.  The surface structure of WDs in CVs is determined by the interplay between accretion and nova explosions which occur through thermonuclear runaways in the accreted material. Measurement of the mass of the accreted layers is of interest for classical nova models, and for assessing the contribution made by novae to the ISM \citep{Gehrz98}. The amount of hydrogen in the accreted envelope versus the accretion rate would show how long it has been since the last nova eruption, something that is otherwise very difficult to measure. Asteroseismological studies have the potential to lead to very precise parameter estimates. The analysis of pulsations in GW Lib by \citet{Townsley04} suggested that the mass of the WD can be constrained to within $3$\%, a level of precision very difficult to observe in the field of CVs, while the mass of the accreted layer can be tied down to $\sim$$20$\%. Masses as precise as this allow the issue of the long-term evolution of WD masses in CVs to be addressed, which is central to Type Ia supernova models. It is unknown as to whether the WDs gain mass as they accrete, or whether they are eroded by nova explosions \citep{Gaensicke00,Littlefair08}. The analysis of GW Lib by \citet{Townsley04} was limited due to only three independent modes being known, making it difficult to identify which normal mode corresponds to which frequency. \citet{Townsley04} assumed that the modes in GW Lib are $l = 1$ spherical harmonics. Asteroseismological mode identification is normally constrained by particular frequency differences and ratios, multiplet structures, and direct period matching. The identification problem is better constrained if a large number of frequencies can be identifed. \citet{Townsley04} predicted additional mode periods that would require more sensitive observations in order to be observed. 

Another important feature of the CV pulsators is that they are subject to irregular heating events from accretion-driven outbursts. GW Lib was caught on the rise to outburst on the 12 April 2007: the first outburst observed in this source since the one which led to its discovery in 1983, and the first outburst to be observed in any CV known to contain a pulsating WD. Events such as this one allow us to study the interplay between the WD temperature and the pulsations. CV pulsators all exist within a well constrained region of parameter space: in order for the pulsations to be detectable they must be low luminosity systems in which the WD dominates the flux. However, some spectral fits to CV pulsators suggest a WD surface temperature that is too high to fall in the instability strip for non-accreting ZZ~Ceti stars \citep{Szkody02,Szkody07a}. This discrepancy may be a result of abundance and atmospheric temperature differences in the accreting systems \citep{Arras06}, due to accretion of He-rich material \citep{Gaensicke03}. Unlike the instability strip for isolated ZZ~Ceti WDs, the empirical boundaries of the CV instability region are yet to be determined with precision \citep{Gianninas06b}. 

In this paper we report high time-resolution observations of GW~Lib and SDSS 1610 taken in May 2005. Our observations were simultaneous in multiple bands and of a higher sensitivity than previous studies. We therefore sought to determine additional pulsation modes to those already known and to investigate the colour-dependence of these pulsations, in order to provide more reliable determinations of the system parameters. We also report high time-resolution observations of GW Lib taken in 2007 in order to examine the effect of heating due to the outburst on the WD pulsations. We aimed to determine if the heating of the WD had affected the known modes (perhaps even suppressing them entirely due to the WD being pushed above the CV instability region) and we sought also to determine if new periodicities were visible in this source, due the WD having been moved into higher $T_{eff}$ instabilities caused by ionisation of helium. We continued to monitor GW Lib through 2008 so as to examine the evolution of the pulsations as the WD cools.

In Section \ref{sec:obs} of this paper we detail our observations and data reduction. In Sections \ref{sec:gwlib_may05} and \ref{sec:gwlib_jun07} we present our results for GW Lib (before and after outburst, respectively). In Section  \ref{sec:1610_may05} we present our results for SDSS 1610. In these three sections we give lightcurves and variability amplitude vs frequency spectrograms (which we will refer to as `amplitude spectra' in this paper). In Section \ref{sec:disc} we examine these results further and discuss their implications for the physical nature of the WDs in these two systems.

\section{OBSERVATIONS}
\label{sec:obs}

\begin{table*} 
\caption{Log of the observations.}
\label{tab:obs} 
\begin{tabular}{lllllllll} 
	    	&   	    	&\multicolumn{2}{c}{UT}     	&		&Avg. exposure			&&\\
Source 	    	&Start date  	&start	    	&end    	&Filter(s)	&time (secs)	&Binning &Conditions\\
\hline
\multicolumn{6}{c}{\it VLT + ULTRACAM, May 2005}\\
GW Lib	    	&07 May     	&03:09  	&09:54  	&$u'g'r'$	&4	&$1\times 1$	&Clear, seeing $0.9$ -- $1.5''$\\
    	    	&08 May     	&05:08  	&09:46  	&$u'g'r'$	&4	&$1\times 1$	&Clear, seeing $1.0$ -- $1.5''$\\\
    	    	&15 May	    	&04:50  	&06:00  	&$u'g'r'$	&2	&$1\times 1$	&Light cloud, seeing $\sim$$0.6''$\\\    	
SDSS J1610 	&09 May	    	&05:32  	&09:54  	&$u'g'r'$	&10	&$2\times 2$	&Light cloud, seeing $0.8$ -- $1.5''$\\
    	    	&10 May	    	&04:56  	&10:09  	&$u'g'r'$	&10	&$2\times 2$	&Clear, seeing $0.5$ -- $0.6''$\\
\hline
\multicolumn{6}{c}{\it SAAO 1.9m + UCT CCD photometer, May 2005}\\
GW Lib	    	&07 May        	&19:20 	    	&02:01 		&none		&15	&$6\times 6$ to $3\times 3$	&Light cloud, seeing decreasing from $5$ to $2''$\\
    	    	&08 May        	&18:38 	    	&01:59 		&none		&15	&$5\times 5$	&Clear, seeing $\sim$$4$''\\   	
    	    	&12 May	    	&20:54 	    	&00:01 		&none		&15	&$6\times 6$	&Clear, seeing $\sim$$4$''\\   	
SDSS J1610 	&09 May	    	&19:38 	    	&02:34 		&none		&60	&$5\times 5$	&Clear, seeing $4$''\\
    	    	&10 May        	&22:19 	    	&01:25 		&none		&60	&$5\times 5$	&Heavy cloud towards end of run.\\   	
\hline
\multicolumn{6}{c}{\it VLT + ULTRACAM, June 2007}\\
GW Lib		&13 June    	&22:41	    	&00:27      	&$u'g'i'$	&3	&$1\times 1$	&Heavy cloud, seeing $\sim$$1.5''$\\
		&14 June	&22:54	    	&01:10		&$u'g'i'$	&0.65	&$1\times 1$	&Heavy cloud, seeing $\sim$$1''$\\
		&16 June	&01:35	    	&04:37		&$u'g'r'$	&0.65	&$1\times 1$	&Some light cloud, seeing $0.6$--$0.7''$\\
		&18 June	&22:58	    	&01:27		&$u'g'r'$	&0.65	&$1\times 1$	&Clear, seeing $0.7$ -- $1.5''$\\
\hline
\multicolumn{6}{c}{\it WHT + API, July 2007}\\
GW Lib		&22 July	&20:50	    	&23:09		&$B$		&30	&$1\times 1$	&Clear, seeing $\sim$$1.6''$\\
		&23 July	&20:56	    	&22:58		&$B$		&30	&$1\times 1$	&Clear, seeing $1.2$ -- $1.5''$\\
\hline
\multicolumn{6}{c}{\it LT + RATCam, March -- June 2008}\\
GW Lib		&08 March	&03:00	    	&05:06		&$g'$		&30	&$2\times 2$	&Heavy cloud, seeing $2$ -- $3''$\\
		&11 March	&02:51	    	&04:56		&$g'$		&30	&$2\times 2$	&Light cloud, seeing decreasing from $5$ to $2''$\\
		&16 March	&02:50	    	&04:56		&$g'$		&30	&$2\times 2$	&Generally quite clear, seeing $< 2''$\\
		&19 March	&02:19	    	&04:24		&$g'$		&30	&$2\times 2$	&Clear, $2''$ seeing\\
		&20 March	&02:31	    	&04:36		&$g'$		&30	&$2\times 2$	&Clear, $2$ -- $3''$ seeing\\
		&30 March	&01:31	    	&03:36		&$g'$		&30	&$2\times 2$	&Light cloud, seeing $2$ -- $3''$\\
		&31 March	&01:39	    	&03:44		&$g'$		&30	&$2\times 2$	&Light cloud, seeing $3$ -- $5''$\\
		&12 April	&00:38	    	&02:43		&$g'$		&30	&$2\times 2$	&Moderate cloud, seeing $1$ -- $2''$\\
		&29 April	&23:57	    	&02:02		&$g'$		&30	&$2\times 2$	&Moderate cloud, seeing $2$ -- $5''$\\
		&11 May		&22:59	    	&01:04		&$g'$		&30	&$2\times 2$	&Light cloud, seeing $1$ -- $2''$\\
		&01 June	&21:39	    	&23:44		&$g'$		&30	&$2\times 2$	&Light cloud, seeing $1$ -- $2''$\\
		&02 June	&21:40	    	&23:45		&$g'$		&30	&$2\times 2$	&Light cloud, seeing $1$ -- $3''$\\
		&21 June	&21:26	    	&23:30		&$g'$		&30	&$2\times 2$	&Clear, seeing $\sim$$1''$\\

\end{tabular}
\end{table*}

A complete log of the observations is given in Table \ref{tab:obs}.

The high speed CCD camera ULTRACAM \citep{Dhillon07} was mounted on the European Southern Observatory (ESO) VLT UT3 ({\it Melipal}) in May 2005 as a visiting instrument. Two nights of this run were dedicated to observations of GW Lib and SDSS 1610. ULTRACAM is a triple beam camera and the observations of both targets were made using the SDSS $u'$, $g'$ and $r'$ filters. The dead time between frames for all the VLT+ULTRACAM data was $25$ms. GW Lib was observed on $7$th and $8$th May. There is a $\sim$$20$ min gap in the $7$th May data due to the target passing through the zenith blind spot of the telescope. We took an additional, shorter observation of this target $\sim$$1$ week later on the $15$th of May. The data were unbinned and the CCD was windowed in order to achieve the required exposure time. SDSS 1610 was observed on the $9$th and $10$th of May. This source is much fainter than GW Lib ($V$ magnitude $\sim$$19$, \citealt{Woudt04}), and so a longer exposure time was used and a CCD window was not required. A binning factor of $2 \times 2$ was used for the VLT+ULTRACAM observations in order to improve the count rate in the $u'$-band. 

Additional, complementary observations were made in 2005 with the University of Cape Town (UCT) CCD photometer \citep{ODon95}, mounted on the $1.9$ metre telescope at the South African Astronomical Observatory (SAAO). This instrument was used in frame-transfer mode and the observations were made in white light in order to maximise the count rate. GW Lib was observed on the $7$th, $8$th and $12$th of May, and SDSS 1610 was observed on the $9$th and $10$th May.

\begin{figure}
\centering
\includegraphics[angle=270,width=1.0\columnwidth]{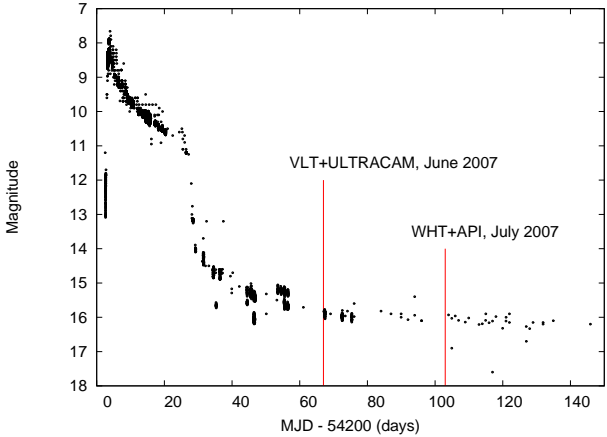}
\caption{$V$-band lightcurve for GWLib, taken between April and August 2007. These data are taken from 
 the website of the American Association of Variable Star Observers (http://www.aavso.org/), and show the outburst and subsequent decline. We mark the times of our observations with VLT+ULTRACAM and WHT+API.} \label{fig:gwlib_aavso} \end{figure}

GW Lib went into outburst in April 2007. In Figure \ref{fig:gwlib_aavso} we plot five months of amateur observations of this source, following the initial outburst. We see in this figure that the $V$-band magnitude of the source rapidly rose to $\sim$$8$ as the disc moved into a high state. the luminosity of the source then declines over $\sim 20$ days as the amount of matter in the disc decreases. This is followed by a very rapid decline as the disc returns to a low state. From $\sim 30$ days after the initial rise to outburst, we see a much more gradual decline in luminosity, leading us to believe that emission from the heated WD began to dominate. On $23$rd May the spectrum of GW Lib showed broad absorption lines (Steeghs, private communication), confirming this belief. In June 2007 we were awarded discretionary time with VLT+ULTRACAM with which to study the effects of the April 2007 outburst of GW Lib. The $V$-band luminosity of the source at this point was estimated to be $\sim$$16$ -- still more than a magnitude brighter than before the outburst. We were therefore able to use a much shorter exposure time compared to our May 2005 observations. The data were unbinned and we used the same CCD window as for the May 2005 observation. For the first two nights, we used the SDSS $i'$ filter in place of $r'$, for scheduling reasons.

We observed GW Lib again on $22$nd and $23$rd July 2007 with the Auxilary Port Imager (API) on the $4.2$m William Herschel Telescope (WHT).  We used the Harris $B$-band filter, and an exposure time of $30$ seconds. In $2008$ we began a monitoring program for GW Lib, using RATCam on the Liverpool Telescope (LT; \citealt{Steele04}). In this paper we present twelve two-hour blocks of data taken between March and June. We used the SDSS $g'$ filter and a $2 \times 2$ binning. The exposures were $30$ seconds in length, with a $\sim$$10$ s dead time.

All of these data were reduced with aperture photometry using the ULTRACAM pipeline software, with debiassing, flatfielding and sky background subtraction performed in the standard way. The source flux was determined using a variable aperture (whereby the radius of the aperture is scaled according to the FWHM). Variation in observing conditions were accounted for by dividing the source lightcurve by the lightcurve of a comparision star. The stability of this comparison star was checked against other stars in the field. For the ULTRACAM data we determined atmospheric absorption coefficients in the $u'$, $g'$ and $r'$ bands and subsequently determined the absolute flux of our targets using observations of standard stars taken in evening twilight. We use this calibration for our determinations of the apparent magnitudes of the two sources, although we present all lightcurves in flux units normalised to unity. Using our absorption coefficients we extrapolate all apparent magnitudes to an airmass of $0$. The systematic error introduced by our flux calibration was  $< 0.1$ mag in all bands.

\section{GW Lib: Pulsations in quiescence}
\label{sec:gwlib_may05}

\begin{figure*}
\centering
\hspace*{1.5cm}\includegraphics[angle=270,width=1.0\textwidth]{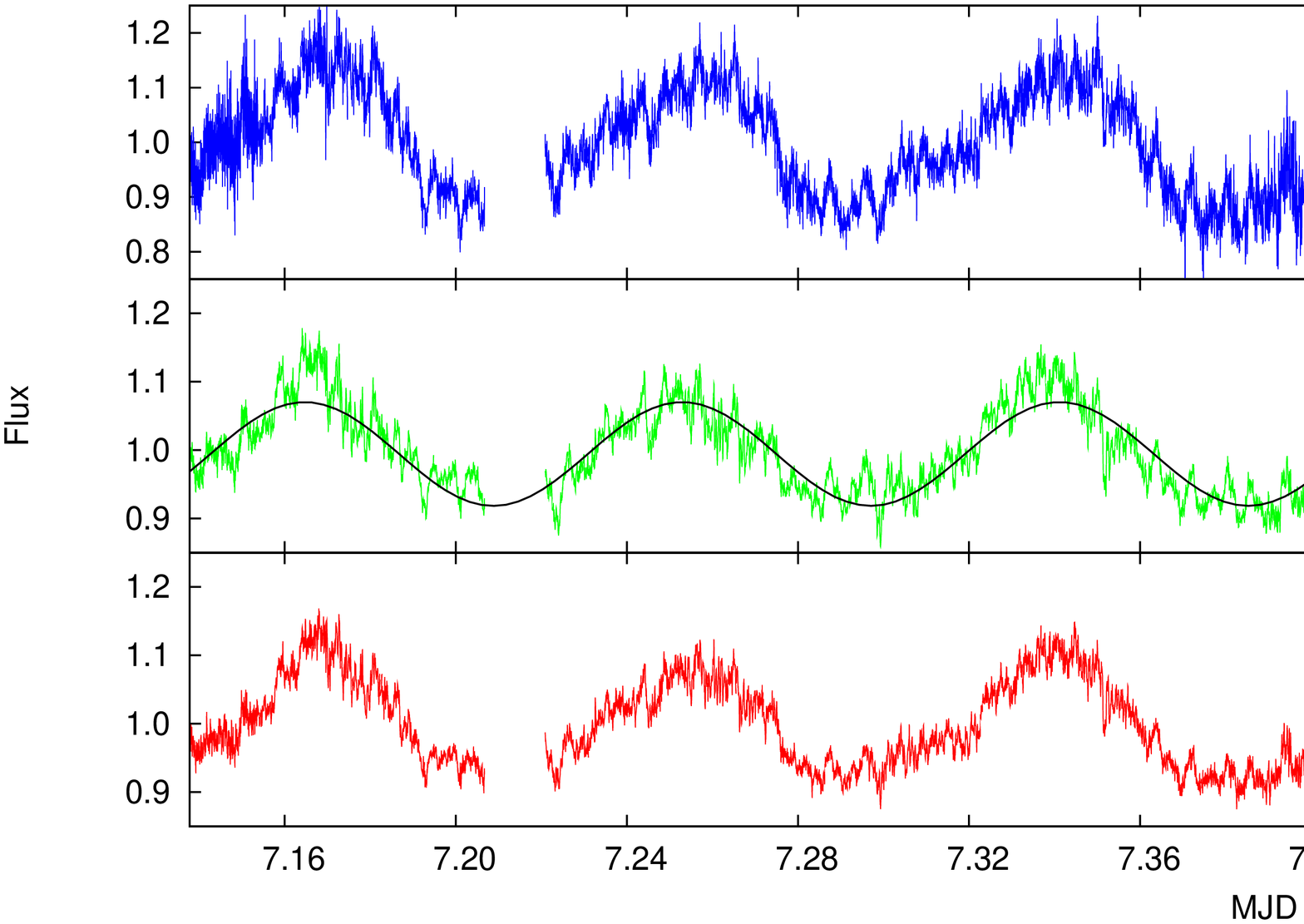}

\hfill

\includegraphics[angle=270,width=1.0\textwidth]{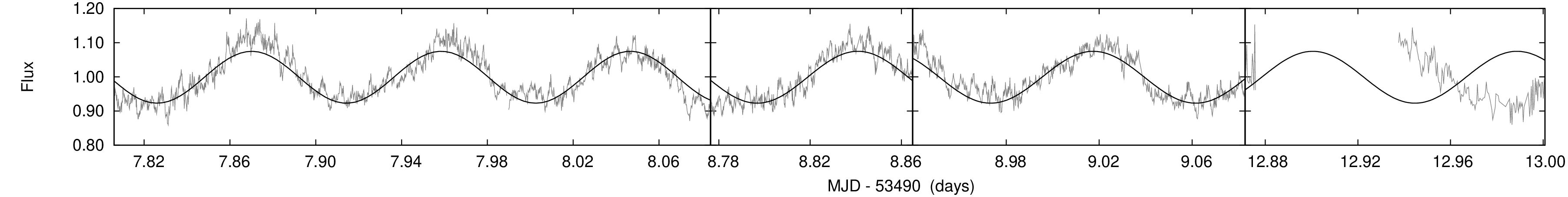}
\caption{Lightcurves for GWLib, taken during quiescence in May 2005. The top plot shows the data taken with VLT + ULTRACAM in the $u'$ (blue, top), $g'$ (green, middle) and $r'$ (red, bottom) filters, taken on the $7$th, $8$th and $15$th May 2005. The bottom plot shows  data taken with the SAAO 1.9m + UCT photometer, on the $7$th, $8$th and $12$th May. We use a flux scale with the mean level normalised to one for both plots. In the $g'$ and SAAO plots, we overplot a fit to the $\sim$$2.1$h modulation, using the model parameters given in Section \ref{sec:long_period}.} \label{fig:gwlib_lc_may05} \end{figure*}

In this section we examine the GW Lib observations taken in May 2005, during which the CV was in its quiescent state. In Figure \ref{fig:gwlib_lc_may05} we show lightcurves of the reduced data. The long period first seen four years prior to these observations by \citet{Woudt02} is apparent, and overlaid on this is significant variation on shorter timescales. Note that this flickering is not instrumental noise: it is intrinsic variation in the source itself. The MJD times given here and in all subsequent plots are on the barycentric dynamical timescale. We find the mean apparent magnitude of GW Lib at this time to be $16.95$ in $r'$, $16.78$ in $g'$ and $17.01$ in $u'$, with amplitudes of $\sim 0.12$, $0.08$ and $0.09$ in $u'$, $g'$ and $r'$ respectively as a result of the long period.

\subsection{Determination of the long period}
\label{sec:long_period}

Before determining the pulsation modes, we first examine the long period. In order to determine the parameters of this modulation, we defined a four-parameter sine function of the form $a\sin \bigl({2\pi} (t-T_0)/P\bigr) + d$. We attempted to fit this model to each night of data separately, as well as the combined dataset. This model provides a good fit, but it can be seen in Figure \ref{fig:gwlib_lc_may05} that the variation is something of a departure from a sinusoid, particularly in $u'$, in which the data appear to have a somewhat saw-toothed shape. However, a fit to a sinusoid is sufficient for determination of the phase and period of this modulation.

We find a consistently good fit between the model and the data when we fit each night of data separately. However, we find that our best-fit parameters are not consistent from night to night. The fitted phase varies by up to $0.072$ and the period takes values of between $2.08$ and $2.13$h. This suggests that the variation is not constant in phase and/or period, although the amplitude of the modulation remains approximately constant. We confirm this when we attempt to fit the entire dataset -- a good fit cannot be found for any constant phase/period model.

We illustrate the changing phase/period of the modulation by plotting a constant phase/period model over the $g'$-band data in Figure \ref{fig:gwlib_lc_may05}. We find that the combined $7$th -- $8$th May data is well fitted by a model with a frequency of $11.335 \pm 0.001$ cycles/day ($P = 2.117$h) and a zero phase of  $T_0 = 53497.26852(8)$ days. The amplitudes in $u'$, $g'$ and $r'$ are $0.12$, $0.08$ and $0.09$ magnitudes respectively. However, it can be seen in the Figure that this model with these parameters is out of phase with the data taken on the $12$th and $15$th of May. The data for these nights are best fitted when a shorter period $P = 2.108$h and a zero phase $T_0 = 53497.338$ are used, but a model with these parameters fits poorly with the $7$th -- $8$th May data.

To summarise, we observe that the $\sim$$2.1$h variation first reported by \citet{Woudt02} is persistent on timescales of years. This phenomenon is quasi-periodic, with a period which varies by minutes over timescales of a few days. We discuss the possible causes of this variation in Section \ref{sec:lp}.

\subsection{Amplitude spectra}
\label{sec:gwlib_ampspec}

\begin{figure*}
\centering
\includegraphics[angle=270,width=1.0\textwidth]{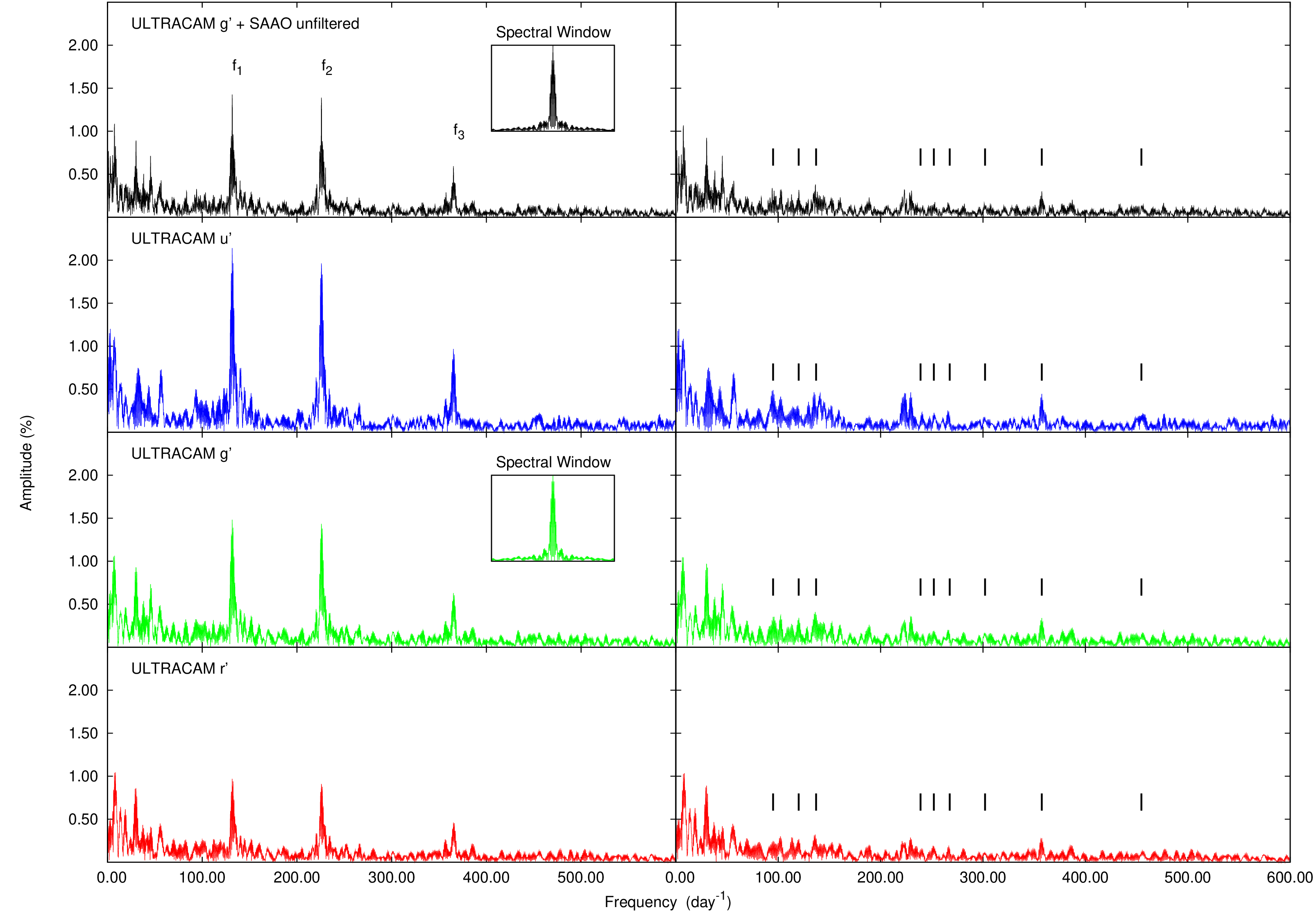}
\hfill
\caption{Amplitude spectra for the May 2005 observations of GW Lib.
 On the left we plot the unwhitened spectra, on the right we plot the same data with the $f_1$, $f_2$ and $f_3$ modes subtracted. Both the left and the right plots have been prewhitened to remove the long period modulation and the majority of the low frequency flickering. The top plots use the VLT+ULTRACAM $g'$ band data combined with the SAAO 1.9m + UCT photometer white-light data. The blue, green and red plots use just the VLT+ULTRACAM $u'$, $g'$ and $r'$ data respectively. Note that the spectral window for the VLT+ULTRACAM $g'$ data is also applicable to the $u'$ and $r'$ plots. In the right hand plot we mark the marginal signals listed in Table \ref{tab:gwlib_periods}.} \label{fig:gwlib_ampspec_may05} \end{figure*}

\begin{figure}
\centering
\includegraphics[angle=270,width=1.0\columnwidth]{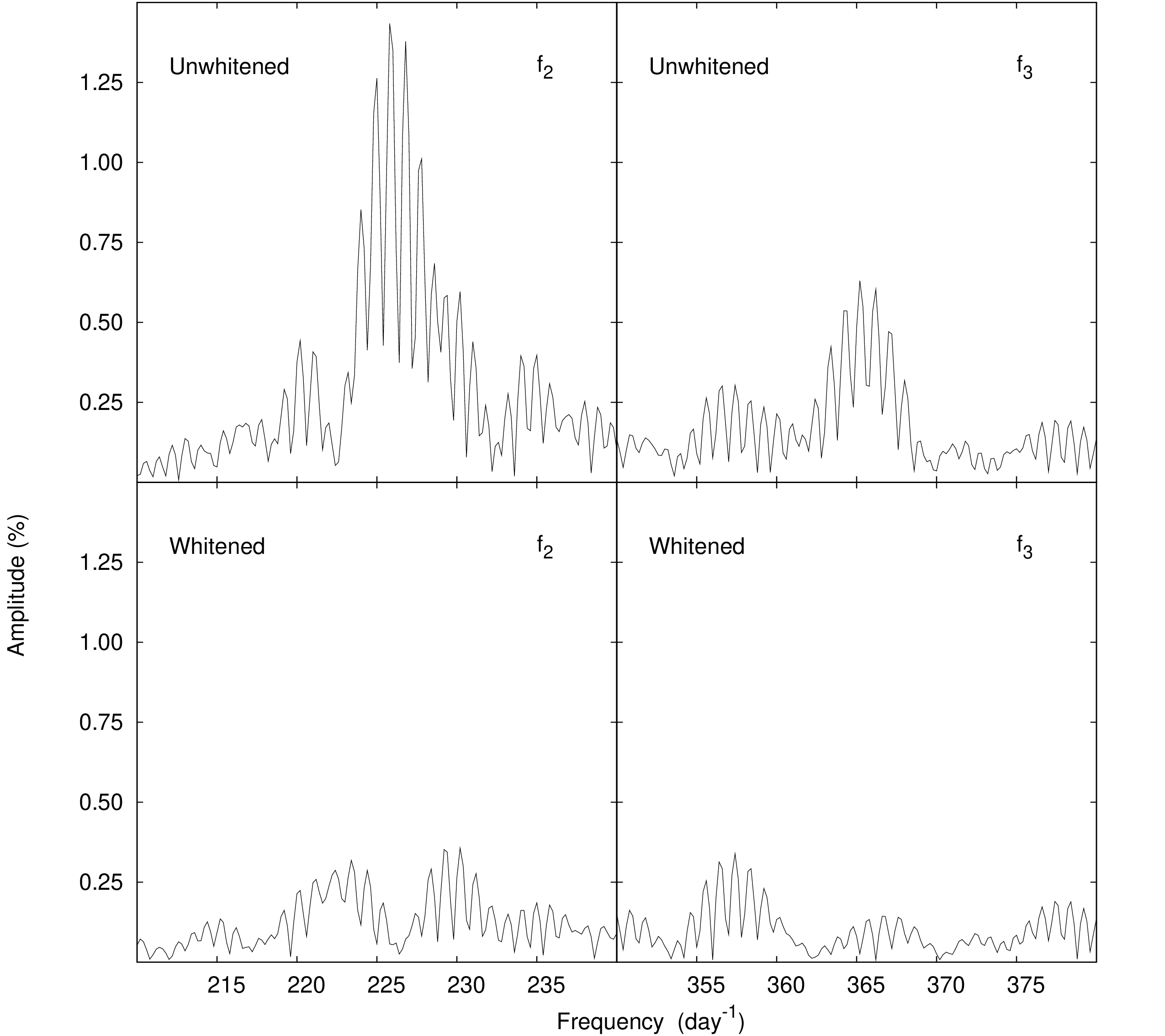}
\hfill
\caption{We plot here the $g'$-band amplitude spectra for the May 2005 observations of GW Lib around the positions of the $f_2$ and $f_3$ modes (taken from the complete spectra given in Figure \ref{fig:gwlib_ampspec_may05}). In the top panels we plot the unwhitened spectra, and in the bottom panels we plot the same data with the $f_2$ and $f_3$ modes subtracted. In the case of the $f_2$ mode (left) the subtraction leaves two residual signals which are very close in frequency to the mode, and are very likely associated with it. In the case of the $f_3$ mode there is a second signal close to the mode, but the seperation is larger and these two signals are more likely to be unrelated.} \label{fig:whitening} \end{figure}

\begin{table*} 
\caption{Main periods in GW Lib during quiescence. The three main periodicities were first identified in \citet{vZyl00}.}
\label{tab:gwlib_periods} 
\begin{tabular}{r@{\,$\pm$\,}ll@{\,$\pm$\,}ll@{\,$\pm$\,}ll@{\,$\pm$\,}ll}
\multicolumn{2}{c}{Frequency}    	&\multicolumn{6}{c}{Amplitudes (\%)}		&\\
\multicolumn{2}{c}{(cycles/day)}	&\multicolumn{2}{c}{$u'$}	&\multicolumn{2}{c}{$g'$}	&\multicolumn{2}{c}{$r'$}		&ID\\
\hline
\multicolumn{9}{c}{\it Main perodicities}\\
$131.639$	&$0.005$		&$2.191$	&$0.036$	&$1.582$	&$0.028$	&$1.042$	&$0.023$		&$f_1$\\
$225.862$	&$0.006$		&$1.916$	&$0.038$	&$1.396$	&$0.030$	&$0.893$	&$0.025$		&$f_2$\\
$365.237$	&$0.013$		&$0.991$	&$0.036$	&$0.698$	&$0.029$	&$0.528$	&$0.025$		&$f_3$\\
\hline
\multicolumn{9}{c}{\it Marginal detections}\\
$95.041$	&$0.094$		&$0.474$	&$0.163$	&$0.407$	&$0.077$	&$0.314$	&$0.068$		&\\
$120.183$	&$0.032$		&$0.349$	&$0.038$	&$0.339$	&$0.028$	&$0.287$	&$0.025$		&\\
$137.248$	&$0.023$		&$0.414$	&$0.037$	&$0.432$	&$0.029$	&$0.316$	&$0.025$		&\\
$239.191$	&$0.050$		&$0.196$	&$0.042$	&$0.167$	&$0.028$	&$0.184$	&$0.024$		&\\
$252.069$	&$0.047$		&$0.231$	&$0.039$	&$0.192$	&$0.028$	&$0.196$	&$0.024$		&\\
$267.522$	&$0.052$		&$0.143$	&$0.039$	&$0.167$	&$0.028$	&$0.162$	&$0.025$		&\\
$302.036$	&$0.172$		&$0.188$	&$0.052$	&$0.165$	&$0.024$	&$0.117$	&$0.023$		&\\
$357.423$	&$0.025$		&$0.390$	&$0.037$	&$0.376$	&$0.028$	&$0.321$	&$0.025$		&$f_1 + f_2$?\\
$454.632$	&$0.167$		&$0.222$	&$0.038$	&$0.165$	&$0.026$	&$0.130$	&$0.022$		&\\
\hline
\end{tabular}
\end{table*}

We determine the frequencies and amplitudes of the WD pulsations  by fitting a model consisting of a series of sine functions to our data. In order to determine the uncertanties on these fits we generated a large number of datasets, resampled from the original data using the bootstrap method \citep{Efron79,Efron93}. The model is fitted to each one of these datasets, generating an array of frequencies and amplitudes for each of the three modes, from which the mean and the RMS error are determined. These uncertainties are an improvement on the formal errors, since as well as the photon and readout noise they include effects such as scintillation. However, in accreting systems the amplitude spectra shows a large amount of high-amplitude, low frequency signals, predominantly due to accretion driven flickering (and in the case of GW Lib, the long period modulation). When the bootstrap method is used on these data this low frequency power can be spread to high frequencies as a result of the poor window function of the resampled data. In order to compensate for this we whiten our data to remove most of the low frequency signals. We first fit sinusoids to each individual night as described in Section \ref{sec:long_period}, and used the resulting fits to remove most of the long period component and any harmonics. We then fit and subtract a polynomial to the data to remove the low frequency flickering power. We find that the uncertainties from the bootstrap do not reduce any further beyond a $\sim 10$th order polynomial. We compute amplitude spectra from these whitened data, which we plot in the left panels of Figure \ref{fig:gwlib_ampspec_may05}. Using the VLT+ULTRACAM data, we plot separate spectra for the $u'$-, $g'$- and $r'$-band data. We plot also the results combining the SAAO 1.9m+UCT photometer observations with the ULTRACAM $g'$-band data (these datasets are not simultaneous). As well as the results shown here, we also calculated separate spectra for each night of observations, in order to check that any signals we detect are persistent over multiple nights.

When we examine Figure \ref{fig:gwlib_ampspec_may05} we see first of all that there are some periodicities clearly evident in these data, in all bands, with amplitudes of $1$ -- $2$\%. There are three strong signals with frequencies of between $100$ and $400$ cycles/day, and a number of low frequency ($< 50$ cycles/day) signals. We see also many peaks at the $0.1$ -- $0.4$\% amplitude level across the entire frequency range. Much of this is accretion-driven `flickering' in the source luminosity. The amplitude of this flickering tends to be highest in the $u'$-band data, and increases at lower frequencies. For example, if we examine the $g'$ data after removing the three dominant signals, we find the mean amplitude to be $0.10$\% $\pm 0.06$ between $100$ and $300$ cycles/day, and $0.08$\% $\pm 0.04$ between $300$ and $600$ cycles/day. This flickering is the dominant source of `noise' in our data (the poisson noise level can be determined by looking at the amplitude spectra at very high frequencies, and we find it to be at the $\sim 0.01$\% level) and the challenge in interpreting these data involves distinguishing genuine periodic signals from this flickering. 

The source flickering increases significantly at low frequencies, and is most likely the cause of the signals we see at $< 50$ cycles/day. We therefore choose to disregard these signals, which leaves three dominant signals with frequencies of between $100$ and $400$ cycles/day. We began by determining the frequencies and amplitudes of these dominant signals, and we list the results in Table \ref{tab:gwlib_periods}, as well as the uncertainties we determined with the bootstrap method. We find these modes to have frequencies consistent with the $f_1$, $f_2$ and $f_3$ modes originally reported by \citet{vZyl00}. 

We whitened our dataset by removing the $f_1$, $f_2$ and $f_3$ modes in order to find weaker signals in our data. We plot the results in the right panels of Figure \ref{fig:gwlib_ampspec_may05}. We find that the whitening leaves some residual peaks very close to the primary peaks. In the case of the $f_1$ mode there is no signal which stands out compared to the surrounding peaks, but there is some evidence of residual signals around the positions of the $f_2$ and $f_3$ modes. In Figure \ref{fig:whitening} we show amplitude spectra around the positions of these two modes before and after whitening. In the $f_2$ case, there are two signals either side of the main peak. These are very close in frequency ($\sim 2$ -- $3$ cycles/day) to the main peak and so we take these to be associated with the $f_2$ mode and do not consider them further. These signals could be spectral leakage due to modulation of the amplitude of the $f_2$ mode by the accretion disc. If we now examine the amplitude spectra around the $f_3$ mode, we see after whitening there is a signal left in the data with a frequency $\sim 8$ cycles/day lower than the $f_3$ mode. In this case, the separation between the mode and the `residual' peak is sufficently large for us to consider these two peaks to be distinct.

Other than these signals near the $f_2$ and $f_3$ modes, we see in Figure \ref{fig:gwlib_ampspec_may05} that there are no signals other than the main three modes that stand out as being particularly strong in amplitude over the flickering level. There may be periodic signals present at the $0.1$ -- $0.4$\% level, but it is impossible to distinguish them from the source flickering. The criteria by which we try to determine real periodicities in our data is somewhat subjective. We looked for signals which have an amplitude that is greater than the mean level in their vicinity, and which appear to be present on every night. We then determined the frequencies and amplitudes of all of these signals by simultaneously fitting a series of sine functions to the dataset. The uncertainty on each signal was determined using the bootstrap method described above. We eliminated from further consideration any signals for which the determined error on the amplitude was comparable to or greater than the value itself. We are left with a list of nine candidates for periodic signals in our data. We  list these detections in Table \ref{tab:gwlib_periods}, but they should be treated as marginal at best. Some are better candidates than others: the aforementioned signal near the $f_3$ mode seems significant. We note also that this signal has a frequency of $357.424$ cycles/day ($P=242$s), and is within $3 \sigma$ of the position of the $f_1 + f_2$ linear combination. We note also that we see a number of low frequency ($< 100$ cycles/day) signals, none of them consistent with the spectroscopic period reported by \citet{Thorstensen02} (${\nu}_{orb} = 18.75$ cycles/day).

\section{GW Lib: Pulsations after outburst}
\label{sec:gwlib_jun07}

\begin{figure*}
\centering
\includegraphics[angle=270,width=1.0\textwidth]{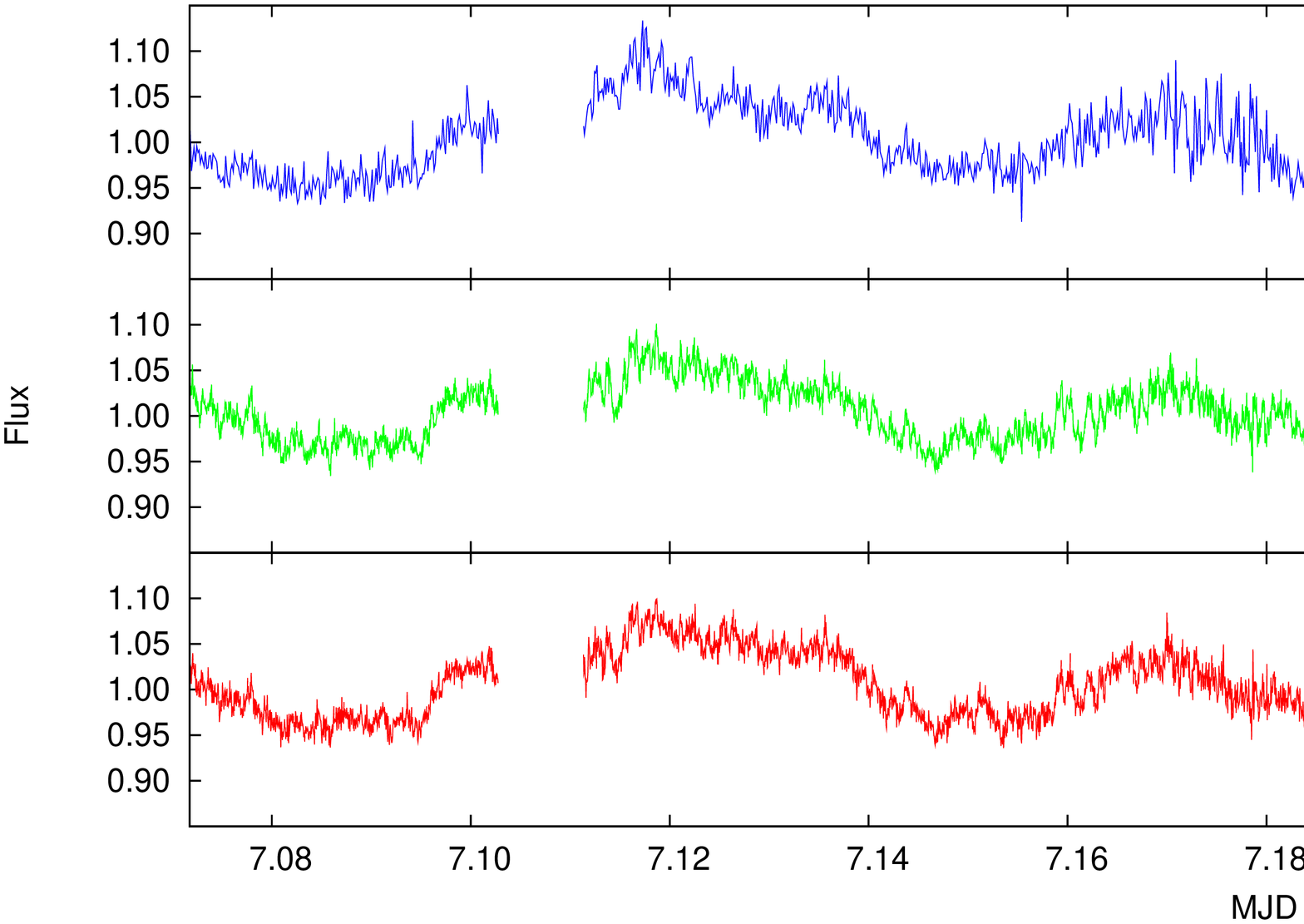}

\hfill

\hspace*{2cm}\includegraphics[angle=270,width=1.0\textwidth]{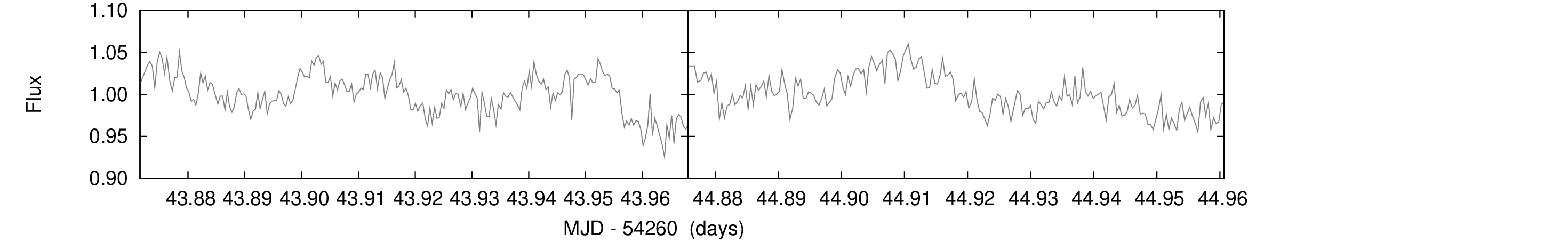}
\caption{Lightcurves for GW Lib, taken a few months after outburst. The top plot shows the data taken with VLT + ULTRACAM in the $u'$ (blue), $g'$ (green) and $r'$ (red) filters in June 2007. The $u'$-band data is binned in order to compensate for poor conditions. The bottom plot shows the data taken with WHT+API in July 2007, with a $g$' magnitude scale.} \label{fig:gwlib_lc_jun07} \end{figure*}

In this Section we examine the GW~Lib data, taken in the aftermath of the April 2007 outburst. We plot the lightcurves of these data in Figures \ref{fig:gwlib_lc_jun07} and \ref{fig:gwlib_lc_lt}.

\subsection{June/July 2007: two -- three months after outburst}

\begin{figure}
\centering
\includegraphics[angle=270,width=1.0\columnwidth]{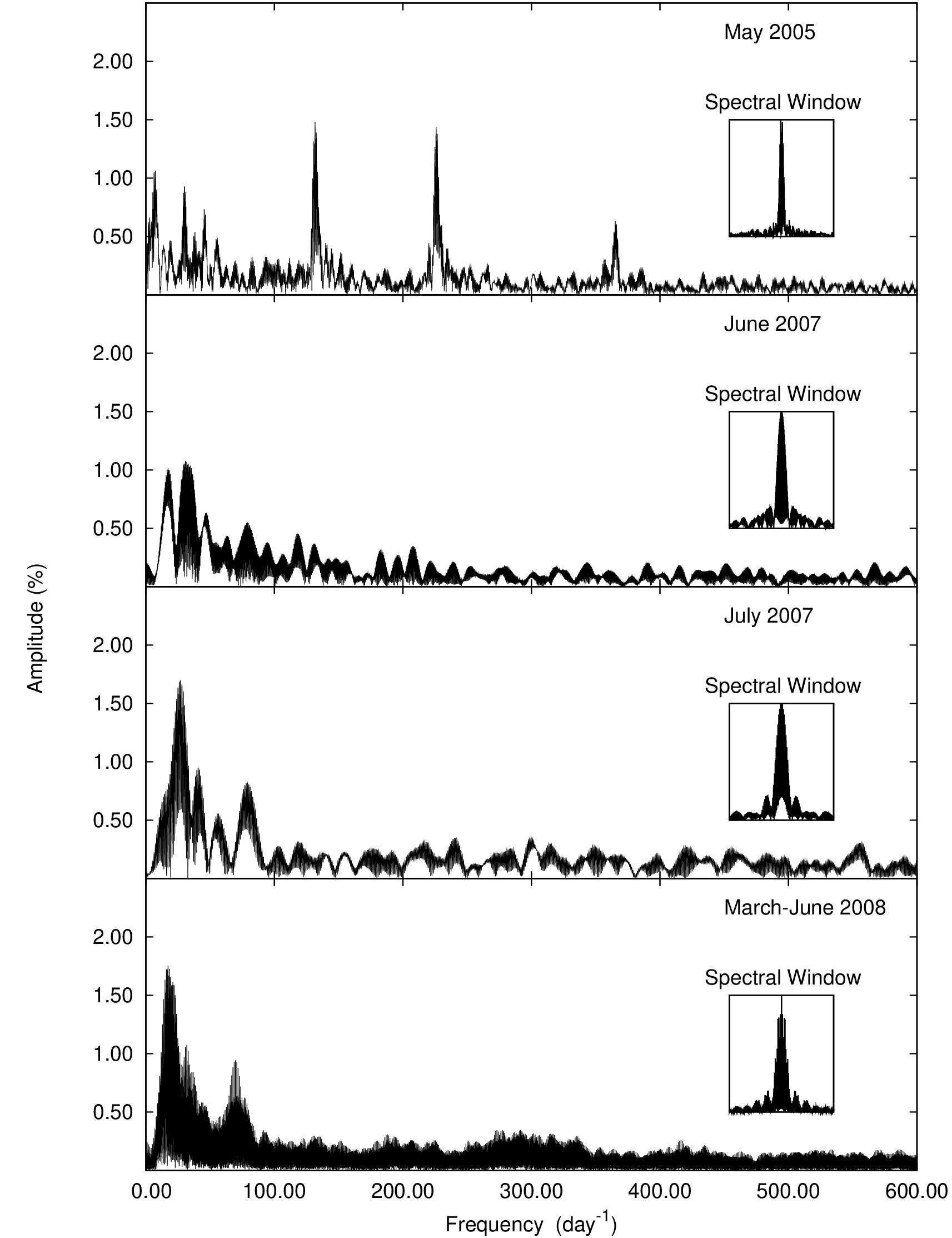}
\hfill
\caption{Amplitude spectra for GW Lib. In the top panel, we plot the quiescent $g'$ data taken in May 2005. In the other three panels, we plot the three epochs of post-outburst data. In the second panel, we plot the $g'$-band amplitude spectrum obtained in June $2007$ with VLT+ULTRACAM. in the third panel, we plot the spectrum obtained in July $2007$ with WHT+API. In the bottom panel, we plot the $g'$-band spectrum of the combined dataset obtained with LT+RATCam.} \label{fig:gwlib_ampspec_jun07} \end{figure}

We begin by discussing the data collected a few months after the outburst (Figure \ref{fig:gwlib_lc_jun07}). For the June $2007$ VLT+ULTRACAM data, we plot only the data taken on the $16$th and $18$th of June, since the other data were seriously affected by poor weather conditions. We plot the complete dataset obtained in July $2007$ with WHT+API. The mean apparent magnitudes of GW Lib are $15.6$, $15.5$ and $15.2$ in $u'$, $g'$ and $r'$ at this time: an increase in brightness of $\sim 1.35$, $1.25$ and $1.81$ magnitudes respectively over the mean values we determined in quiescence. The source is both significantly more luminous and bluer in colour in these months after the outburst. 

There is some large amplitude, long period modulation apparent in these data. This variation is clearly not sinusoidal but does appear to be somewhat periodic. While we cannot fit this modulation with any degree of certainty we find it to have a period close to the spectroscopic period reported by \citet{Thorstensen02}, and so this may be an orbital modulation. Secondly, we note that these lightcurves do not show the coherent, short-period variation indicative of the pulsation of the WD, which was so obvious in the quiescent data (Figure \ref{fig:gwlib_lc_may05}).

We plot the amplitude spectra for the four epochs of GW Lib data in  Figure \ref{fig:gwlib_ampspec_jun07}. We see in this Figure that in June and July $2007$, two to three months after the outburst, we do not detect any of the pulsation modes that were observed in the quiescent data. The luminosity of the source, combined with the very gradual decline we see in Figure \ref{fig:gwlib_aavso}, leads us to believe that the WD makes a larger contribution to the flux in these post-outburst data compared to quiescence, and thus one would expect to detect the modes 
more easily if they persist. The spectra are dominated by low-amplitude flickering at all frequencies. In the VLT+ULTRACAM data, this flickering level is $\sim$$0.2$\%, which is the same as in the equivalent quiescent data. However, we do observe the source flickering to be of a slightly higher amplitude at low frequencies, with an amplitude of $\sim$$0.3$ -- $0.5$\% over the frequency range where we previously observed the $f_1$ and $f_2$ modes. There is no single signal that stands out as being a coherent pulsation mode. In the WHT+API data, we see a couple of peaks with amplitudes $\sim$$0.5$\% at a frequency of between $50$ and $100$ cycles/day. These may just be flickering, or the largest peak may be related to the strong signal we see in the March $2008$ data. (see Section \ref{sec:2008obs}). If this is an early detection of this signal, it is marginal at best.

\subsection{March -- June 2008: $\sim$$1$ year after outburst}
\label{sec:2008obs}

\begin{figure*}
\centering
\includegraphics[angle=270,width=1.0\textwidth]{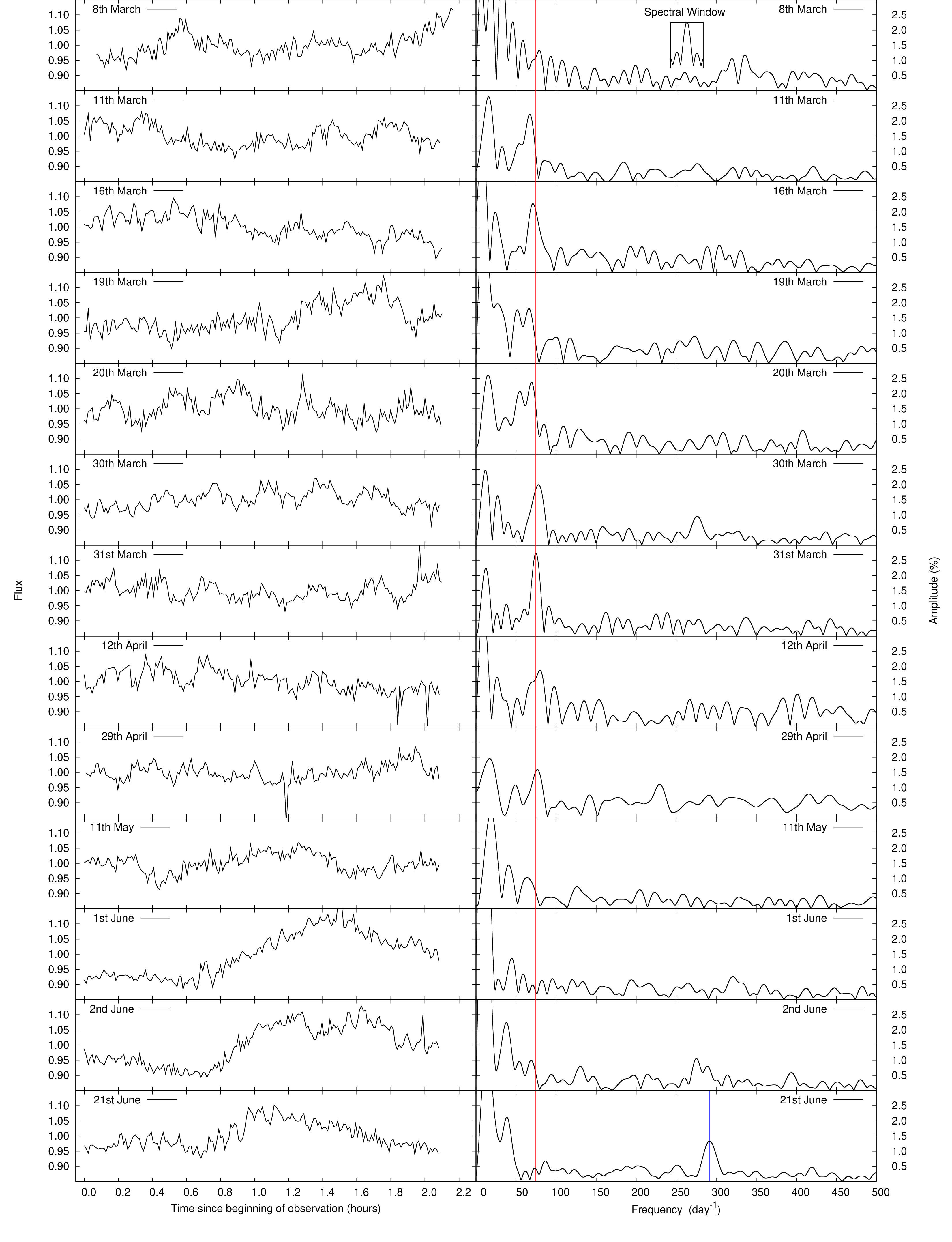}
\caption{The datasets obtained for GW Lib with the Liverpool Telescope + RATCam, taken between eleven and fourteen months after outburst. We plot lightcurves (left) and amplitude spectra (right), using a $g'$ filter. On the amplitude spectra we mark with a red line the position of the $\sim$$75$ cycles/day periodicity observed in late March and early April. In the $21$st June panel we mark with a blue line the position of the  $\sim$$292$ cycles/day periodicity.} \label{fig:gwlib_lc_lt} \end{figure*}

When we flux calibrate the LT+RATCam data we find that eleven months after the outburst, the source is still more than half a magnitude more luminous than during quiescence, with a mean $g'$-band apparent magnitude of $\sim$$16.2$. We plot the data on a relative flux scale in Figure \ref{fig:gwlib_lc_lt}. We see that there is a degree of luminosity variation throughout, and in particular in the June observations we see a variation with an apparent period of the order of $2$ hours. This may be the re-emergence of the long period variation observed in quiescence. 

We do not see the quiescent periodicities in any of these data. However, over the course of these runs we do see a new periodic signal emerge at a frequency of $\sim$$75$ cycles/day. There is evidence for this signal from $11$th March onwards, but in the early March data it appears to drift in frequency. This may be real, or it may be the result of confusion with low-amplitude flickering. This signal is consistent in frequency in the $30$th March -- $29$th April data. Throughout April it declines in amplitude, and it is not apparent in the May/June data. We fit a sine function to the $31$st March data where this signal is most prominent, and find it to have a frequency of $74.86 \pm 0.68$ cycles/day ($P=1154$s) and a $g'$ amplitude of $2.20$\% $\pm 0.18$. This amplitude is greater than that of the periodicities observed in the quiescent datasets.

In the last dataset, taken on the 21st June, there is evidence for another new periodicity in this source. When we fit this signal we find it to have a frequency of $292.05 \pm 1.11$ cycles/day ($P=296$s) and a $g'$ amplitude of $1.25$\% $\pm 0.18$.

\section{SDSS 1610: Pulsations in quiescence}
\label{sec:1610_may05}

\begin{figure*}
\centering
\hspace*{0.5cm}\includegraphics[angle=270,width=1.0\textwidth]{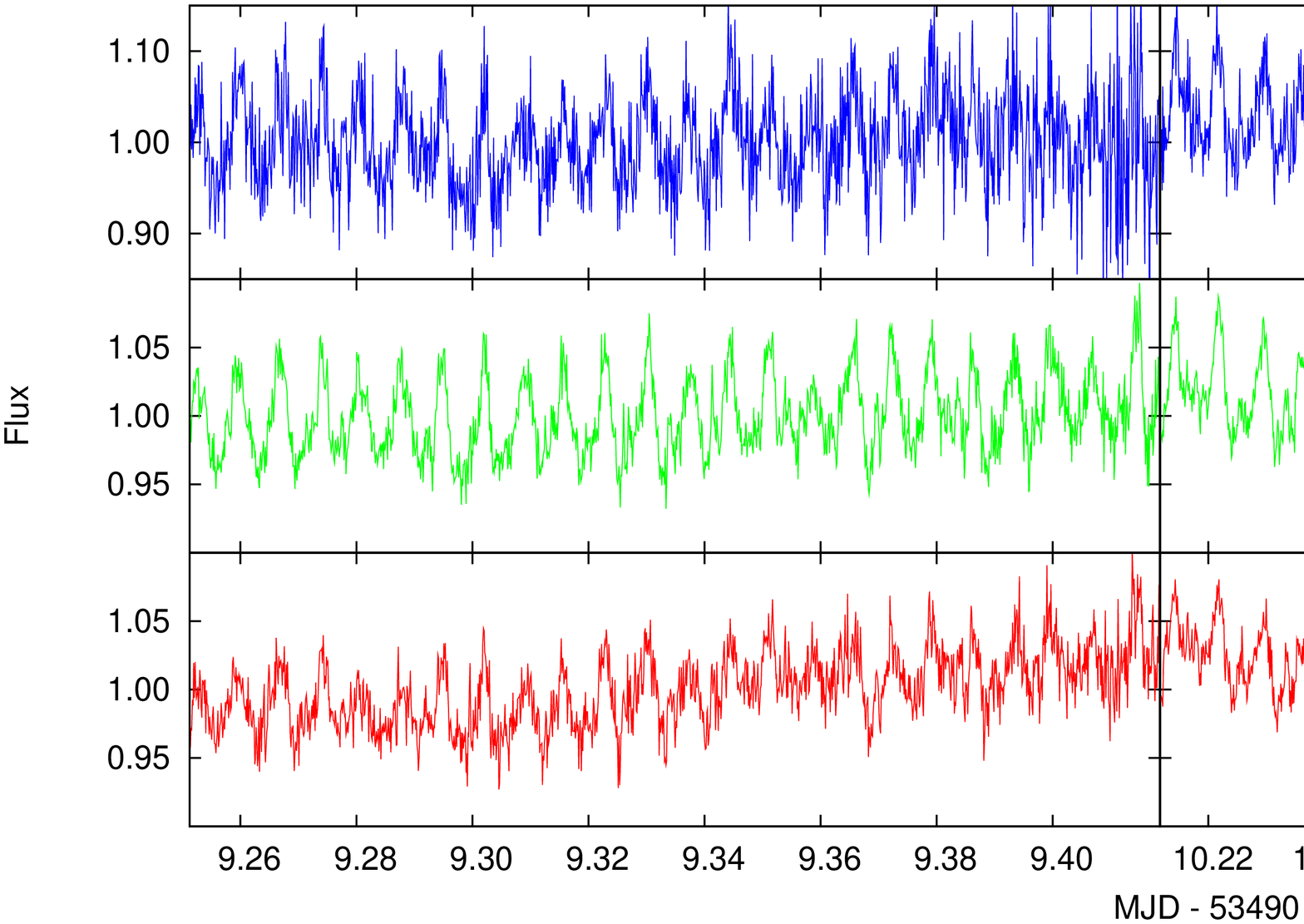}

\hfill

\includegraphics[angle=270,width=1.0\textwidth]{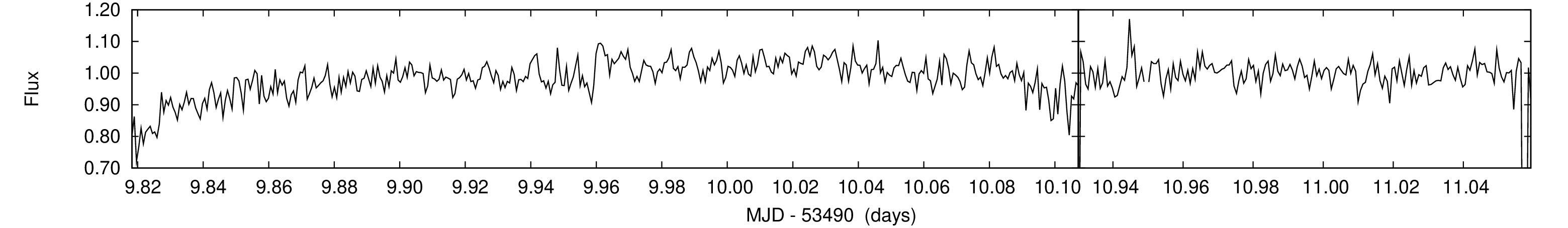}
\caption{Lightcurves for SDSS 1610, taken during quiescence in May 2005. The top plot shows the data taken with VLT + ULTRACAM in the $u'$ (blue), $g'$ (green) and $r'$ (red) filters. The bottom plot shows the data taken with the SAAO 1.9m + UCT photometer. We use a magnitude scale for the VLT+ULTRACAM data, and a flux scale with the mean level normalised to one for the white-light SAAO data.} \label{fig:1610_lc} \end{figure*}

In this Section we examine the SDSS 1610 observations taken in May 2005. We plot the lightcurves in Figure \ref{fig:1610_lc}. The pulsations of the WD can be clearly seen in these data. There is no clear evidence for a long period modulation in this source similar to that observed in GW Lib. The gradual variation in the mean count rate that can be observed in the SAAO observations is due to the changing airmass over the course of these white-light observations. 

\begin{figure}
\centering
\includegraphics[angle=270,width=1.0\columnwidth]{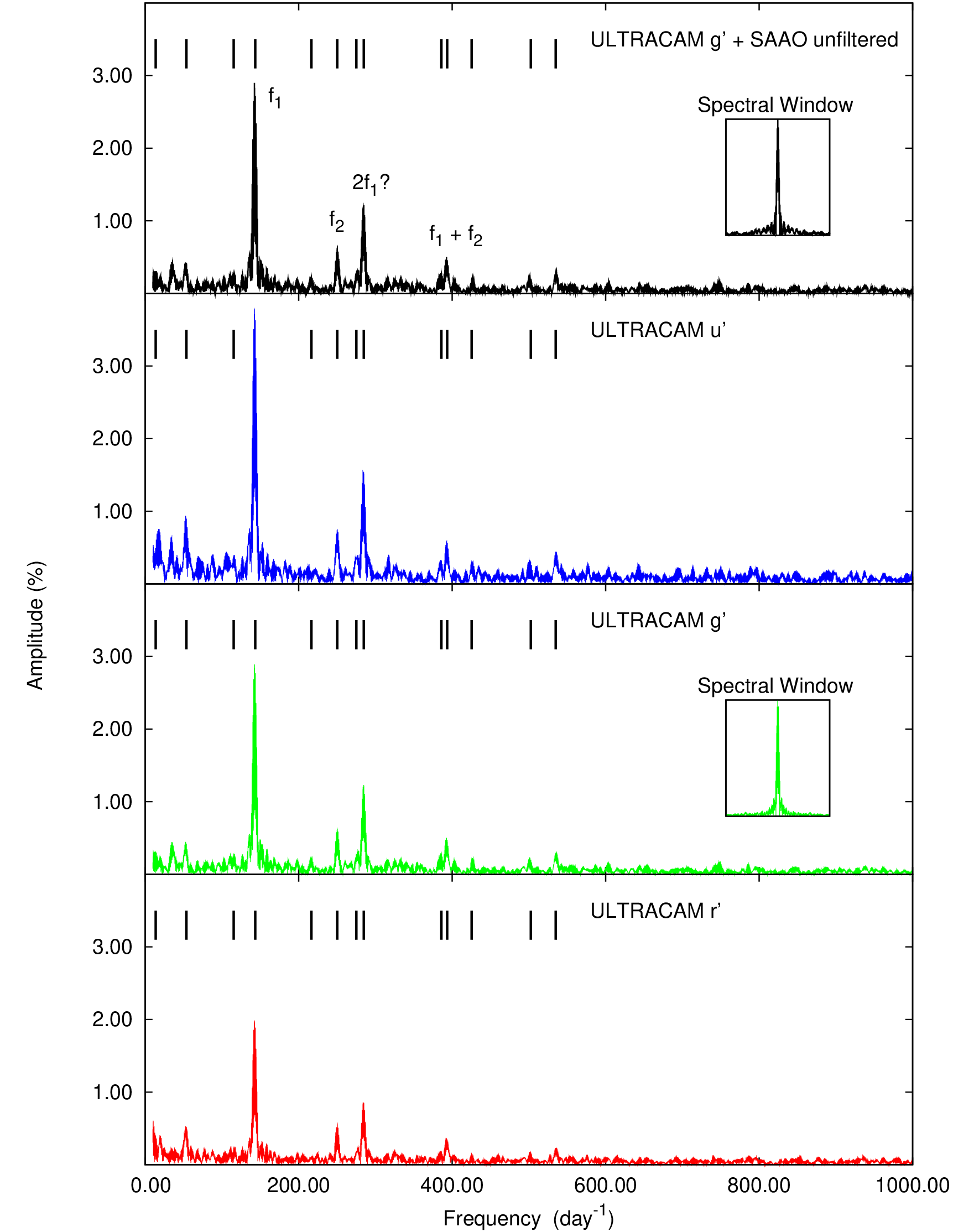}
\hfill
\caption{Amplitude spectra for the May 2005 observations of SDSS 1610. The top plot uses the VLT+ULTRACAM $g'$ band data combined with the SAAO 1.9m + UCT photometer white-light data. The blue, green and red plots use just the VLT+ULTRACAM $u'$, $g'$ and $r'$ data respectively. We mark the signals listed in Table \ref{tab:1610_periods}, with the longer marks showing the main modes and the possible linear combinations.} \label{fig:1610_ampspec} \end{figure}

\begin{table*} 
\caption{Main periods in SDSS 1610. The two primary modes and four combinations were first identified in \citet{Woudt04}.}
\label{tab:1610_periods} 
\begin{tabular}{r@{\,$\pm$\,}ll@{\,$\pm$\,}ll@{\,$\pm$\,}ll@{\,$\pm$\,}ll}
\multicolumn{2}{c}{Frequency}    	&\multicolumn{6}{c}{Amplitudes (\%)}		&\\
\multicolumn{2}{c}{(cycles/day)}	&\multicolumn{2}{c}{$u'$}	&\multicolumn{2}{c}{$g'$}	&\multicolumn{2}{c}{$r'$}	&ID\\
\hline
\multicolumn{9}{c}{\it Main periodicities}\\
$143.401$	&$0.004$		&$3.696$	&$0.070$	&$2.817$	&$0.035$	&$1.913$	&$0.035$	&$f_1$\\
$250.232$	&$0.020$		&$0.746$	&$0.075$	&$0.622$	&$0.038$	&$0.540$	&$0.038$	&$f_2$\\
$284.866$	&$0.010$		&$1.573$	&$0.072$	&$1.251$	&$0.035$	&$0.858$	&$0.036$	&$2 f_1$ ?\\
$393.598$	&$0.030$		&$0.546$	&$0.074$	&$0.484$	&$0.035$	&$0.350$	&$0.035$	&$f_1 + f_2$\\
\hline
\multicolumn{9}{c}{\it Other signals}\\
$13.638$	&$0.023$		&$0.851$	&$0.116$	&$0.537$	&$0.057$	&$0.605$	&$0.048$	&\\
$53.715$	&$0.031$		&$0.873$	&$0.085$	&$0.413$	&$0.049$	&$0.518$	&$0.043$	&\\
$115.382$	&$0.051$		&$0.404$	&$0.146$	&$0.330$	&$0.071$	&$0.281$	&$0.061$	&\\
$216.510$	&$0.065$		&$0.107$	&$0.094$	&$0.243$	&$0.040$	&$0.113$	&$0.039$	&\\
$275.132$	&$0.040$		&$0.366$	&$0.071$	&$0.297$	&$0.035$	&$0.213$	&$0.038$	&\\
$385.973$	&$0.057$		&$0.285$	&$0.076$	&$0.245$	&$0.036$	&$0.162$	&$0.035$	&\\
$425.328$	&$0.055$		&$0.325$	&$0.078$	&$0.214$	&$0.036$	&$0.153$	&$0.039$	&\\
$502.476$	&$0.066$		&$0.265$	&$0.091$	&$0.202$	&$0.037$	&$0.155$	&$0.040$	&\\
$534.951$	&$0.069$		&$0.474$	&$0.073$	&$0.319$	&$0.038$	&$0.244$	&$0.037$	&\\
\end{tabular}
\end{table*}

We compute amplitude spectra from these data as we did with GW Lib. We plot the results in Figure \ref{fig:1610_ampspec}. Using the VLT+ULTRACAM data, we plot separate spectra for the $u'$-, $g'$- and $r'$-band data. We plot also a spectrum that is computed from combining the SAAO 1.9m+UCT photometer observations with the ULTRACAM $g'$-band data. We find the mean apparent magnitudes of this source to be $19.10$, $19.04$ and $19.33$ in $u'$, $g'$ and $r'$ respectively, with a variation of $\sim 0.1$ magnitudes in all bands which is due to the pulsations. 

We list the main periods evident in these data in Table \ref{tab:1610_periods}. We use the same method to determine these frequencies, amplitudes and uncertainties as we did for GW Lib. The strongest signals are the peaks which match the periodicities identified by \citet{Woudt04}. We see the two principal modes $f_1$ and $f_2$ at frequencies of $143.40$ and $250.23$ cycles/day respectively ($P=603$ and $345$s), and at frequencies $284.87$ and $393.60$ cycles/day ($P=303$ and $220$s) we see peaks which were presumed by \citet{Woudt04} to be the $2 f_1$ harmonic and the $f_1 + f_2$ combination. Using our calculated uncertainties we find the peak at $393.60$ cycles/day to be consistent with the position of the $f_1 + f_2$ combination to within $1 \sigma$. However, the $284.87$ cycles/day signal is more than $5 \sigma$ from the expected position and so may be an independent mode. We also identify a number of additional signals with amplitudes of $\sim$$0.2$ -- $0.4$\%. We followed the same procedure for identifying potential low-amplitude signals as was used for GW Lib, but for this source the process was much less subjective since the amplitude of the source flickering is much lower. The signals we identify are of a significantly larger amplitude than neighbouring peaks, but in order to confirm the significance of these signals we compared our amplitude spectra to fake datasets consisting only of gaussian white noise. We were hence able to determine the signal/noise amplitude ratio across the entire frequency range. We use the Breger criterion \citep{Breger93} to distinguish between peaks due to pulsation and noise, and we find all of the signals listed in Table \ref{tab:1610_periods} to satisfy this criterion. We therefore have confidence in these signals being real periodicities in the source.

\citet{Woudt04} saw some evidence in their data for the $2 f_2$ harmonic and $2 f_1 + f_2$ combination. We see signals at $502.476$ and $534.951$ cycles/day ($P=172$ and $162$s) which are close to the expected position of these signals, but fall outside of our calculated uncertainties. We see also a number of high amplitude, low frequency ($< 100$ cycles/day) signals. None of these are consistent with the orbital period ($80.52$ min: \citealt{Woudt04}) and they are most likely due to flickering. As with GW Lib, we note the amplitudes of the pulsation modes is highest in the $u'$ band.

\section{DISCUSSION}
\label{sec:disc}

In this Section we examine the results presented in Sections \ref{sec:gwlib_may05} to \ref{sec:1610_may05}. We divide this discussion into three parts. We begin by examining in more detail the periodicities we find in the two sources when we observed them in quiescence. We then discuss the $\sim$$2.1$h modulation in GW Lib. Finally, we discuss the post-outburst observations of GW Lib.

\subsection{Pulsations in GW Lib and SDSS 1610 during quiescence}

\subsubsection{Pulsations in GW Lib}

In Table \ref{tab:gwlib_periods} we list the main periodicities which we observe in the amplitude spectra of GW Lib. Note that the amplitudes we detect are not the true pulsation amplitudes of the WD: there is considerable accretion luminosity present, which dilutes the pulsation amplitudes. We find the spectrum is dominated by three main peaks. These are the pulsation modes discovered and designated $f_1$, $f_2$ and $f_3$ by \citet{vZyl04}. We find a number of additional signals with amplitudes in the $0.2$ -- $0.4$\% range. One is close to the position of the $f_1 + f_2$ linear combination, and was also reported by \citet{vZyl04}. However as we noted in Section \ref{sec:gwlib_ampspec}, the amplitudes of the other signals are comparable to the flickering in the source and so it is impossible to be certain that these are true periodicities. These detections should be treated as being marginal at best. The remaining signals listed cannot be associated with any of the main modes. 

We note also that \citet{vZyl04} reported a number of signals in their data which they identified as linear combinations. These signals were clustered at frequencies of $\sim$$240$, $340$, $580$ and $905$ cycles/day. We see no evidence for these signals in our data. Additionally, we note that the theoretical model of \citet{Townsley04} predicted a number of additional periodicities which should be apparent in GW Lib. One of these periods is $191$s ($452$ cycles/day), which is close to the signal we identify at $454.632$ cycles/day. The remaining predicted periods do not match any of our findings. 

\subsubsection{Pulsations in SDSS 1610}

In Table \ref{tab:1610_periods} we list the main signals we observe in the amplitude spectra of SDSS 1610. As for GW Lib, we should note that the amplitudes we detect are diluted by the accretion luminosity. \citet{Woudt04} reported two main modes and four linear combinations or harmonics in this source. However, our detections of the signals reported as $2 f_1$, $2 f_2$ and $2 f_1 + f_2$ in \citet{Woudt04} have frequencies which are not formally consistent with those identifications. We note also that \citet{Woudt04} reported a number of signals in their data at frequencies $334$, $596$, $711$, $754$ and $839$ cycles/day. These were marginal detections, each of which only appeared in one run. We found no evidence for these signals in our data.

\subsubsection{Colour dependence of pulsations}

\begin{figure}
\centering
\includegraphics[angle=270,width=1.0\columnwidth]{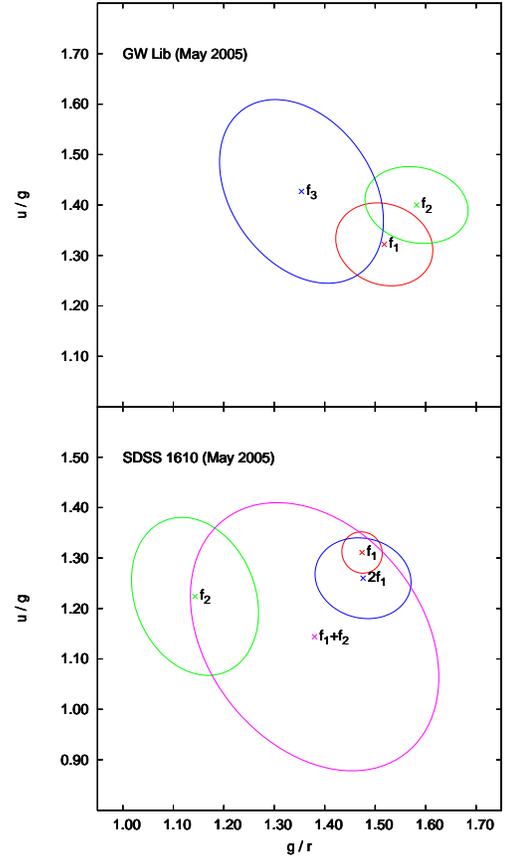}
\hfill
\caption{The colour dependence of the pulsations in the quiescent GW Lib and SDSS1610. We calculate the $g'/r'$ and $u'/g'$ colour ratios for the main signals in each source and plot the resulting $1$-$\sigma$ error contours.} \label{fig:ampplot} \end{figure}

We now investigate the colour dependence of the modes in GW Lib and SDSS 1610. In Figure \ref{fig:ampplot} we plot $g'/r'$ vs. $u'/g'$ for the dominant periodicities in the two sources. We plot also the $1\sigma$ error contour for each signal. These contours are elliptical because we use the $g'$-band flux as a component in both ordinates. We chose the $g'$-band flux since these data have the lowest uncertainty.

  For GW Lib we plot the $f_1$, $f_2$ and $f_3$ modes. For SDSS 1610 we plot $f_1$ and $f_2$, as well as the two high amplitude combinations of these two modes. We see first of all in this plot that both $g'/r'$ and $u'/g'$ are $> 1$ for all signals, indicating that the amplitude of these signals increases at bluer wavelengths in both sources. For GW Lib, we see that the three modes seem to be similar in colour, with all three modes occupying an overlapping region in the parameter space. In the case of SDSS 1610 however, there is a possible discrepancy between the two primary modes. This may be significant. It has been shown that for a given stellar temperature, pressure and geometry, the change in flux as a result of a non-radial pulsation is sensitive to the $l$ number of the pulsation \citep{Watson88}. The fact that the two principal modes in SDSS 1610 have a different colour dependence might suggest that they have different $l$ values.

\subsection{The $2.1$h period in GW Lib}
\label{sec:lp}

The $2.1$h modulation in GW Lib was first observed in $2001$ by \citet{Woudt02}. It was not seen in any previous photometric observations of this source. We observed this modulation in all three bands in $2005$, confirming it to be a persistent feature in the 
lightcurve of GW Lib over several years. Our $2005$ observations of this source were separated by a number of days, and we found that we could not fit this entire dataset with any constant phase/period model. The period of this modulation appears to vary by minutes on timescales of days. There is however no persistent trend since all of the data combined showed a period that was consistent with the finding of \citet{Woudt02} to within a few minutes. It is possible that this apparent variation in phase/period is due to multi-periodicity, but we have insufficient data to properly explore this possibility. 

This variation is difficult to explain. Spectroscopic determination of the orbital period has shown it to be much shorter ($76.78$ min, \citealt{Thorstensen02}). The $2.1$h period therefore cannot be ascribed to an orbital modulation, such as obscuration of the bright spot or an elliptical accretion disc. There are other systems which display a photometric period that is much greater than the spectroscopic period. One example is V2051 Oph, in which a photometric period of $274$ min is found \citep{Warner87}. This is an eclipsing system, and the orbital period has been determined to be $89.9$ min. Another example is FS Aur, in which a long period of $\sim$$3$h was found by \citet{Neustroev02}. This variation was confirmed and found to be persistent by subsequent ULTRACAM observations \citep{Neustroev05}. A further example is V445 And (HS 2331+3905), with spectroscopic and photometric periods at $81.1$ min and $\sim$$3.5$h \citep{Araujo-Betancor04}. One proposed explanation for these sources is an intermediate polar model with a rapidly-rotating and precessing WD \citep{Tovmassian03,Tovmassian07}. However, in the case of GW Lib there is no evidence in the spectra for a strong magnetic field. This model is also inconsistent with our finding of a quasi-periodic nature for the GW Lib modulation.

The fact that this modulation is apparently quasi-periodic in nature suggests it is not directly related to the spin, precession or orbit of the WD. We suggest that it is an accretion-driven phenomenon that most likely originates in the accretion disc. We note also that the July 2007 post-outburst data shows some evidence for a long-period modulation. The pre- and post-outburst modulations may caused by the same phenomena. On one hand, they have consistent amplitudes in absolute flux terms. On the other hand, the period of the post-outburst modulation appears to be close to the spectroscopic period, so this variation may be an orbital modulation due to irradiation/ellipsoidal variations as a result of the changing aspect of the secondary star, or the result of the disc becoming elliptical during outburst, causing superhumps in the lightcurve as seen in the SU~UMa stars. We note also that in the June 2008 LT data, the variation in GW Lib again seems to be dominated by a modulation with a period of $\sim$$2$h. An accurate determination of this period is not possible from these data, since each observation is only $2$h in length.

\subsection{The effects of the outburst in GW Lib}
\label{sec:outburst_disc}

In the post-outburst data we observe GW Lib to be more luminous and bluer in colour than in May 2005. We noted in Section \ref{sec:obs} that the observational data support the conclusion that this increased luminosity is due to the heating of the WD by the April 2007 outburst. The GW Lib outburst is reminiscent of the outbursts in the well-studied system WZ Sge. Studies of the most recent outburst in this system \citep{Patterson02,Kuulkers02,Long03} showed significant heating followed by long-term cooling over a number of years \citep{Sion03,Godon04,Godon06}.

The June/July 2007 lightcurves, taken two -- three months after the outburst (Figure \ref{fig:gwlib_lc_jun07}) show much short-timescale variation, but we do not see the coherent pulsations which are apparent in the quiescent data (Figure \ref{fig:gwlib_lc_may05}). If we examine the amplitude spectra for the June/July 2007 data  (Figure 
\ref{fig:gwlib_ampspec_jun07}), we do not see any well-defined periodicities corresponding to the pulsation modes seen in the quiescent data, or any new periodicities. There are two possible explanations for this. The first is that the heating of the WD has moved it outside of the  CV instability region in the $T_{eff}$ -– $\log g$ plane and the pulsations have been `switched off'. Alternatively, the source may still feature the same coherent pulsations, but they now have an amplitude below the level of the accretion-driven flickering and are therefore undetectable. 

We can estimate a minimum possible percentage amplitude for the pulsations by supposing that the amplitude in absolute flux is unchanged by the heating of the white dwarf. Given our measurements of the source luminosity in May 2005 and June 2007, we find that the minimum amplitude for the $f_1$ and $f_2$ modes in the post-outburst June 2007 data is $\sim$$0.4$\% and the minimum amplitude for the $f_3$ mode is $\sim$$0.2$\%. These amplitudes are approximately the same in all three bands (the change in colour of the source post-outburst compensates for the fact that the amplitudes of the modes is greater at bluer wavelengths during quiescence). Given that the flickering level in our data is generally $\sim$$0.2$\% (approximately the same as in quiescence), we would expect to detect at least the $f_2$ mode, were it present. We do see the flickering increase in amplitude at low frequencies, and at the position of the $f_1$ mode it is  $\sim 0.5$\%, which may be enough to obscure the mode. The presence of this periodicity is unlikely however, given that we would expect to detect the $f_2$ mode at a comparable level. The $0.5$\% amplitude signals at $125$ cycles/day could be a manifestation of the $f_1$ periodicity itself, but since we do not observe a single, coherent peak, we suspect not. 

The fact that the heating of the WD has apparently suppressed or switched off the pulsations is significant. If the pulsations originated deep within the star, then even if the driving mechanism were to be switched-off, one might suspect that the pulsations continue simply due to inertia. The fact that the pulsations have been switched off indicates that in the absence of excitation they are damped on timescales of weeks. This is in line with some current theoretical models: for example \citet{Wu99} predict that $l = 1$ modes in ZZ~Ceti stars with periods comparable to those in GW~Lib will be damped on this timescale.

In the $2008$ data (Figure \ref{fig:gwlib_lc_jun07}), we see that the three known modes are still not present. The luminosity of the source suggests the WD is still significantly hotter at this stage than during quiescence, so this is perhaps not surprising. We do see very clearly in the March/April 2008 data a new signal with a frequency of $\sim$$75$ cycles/day. This is apparently unrelated to any of the known modes. This signal may be a quasi-periodic oscillation originating in the disc, or it may be a WD pulsation. If this signal is associated with the WD, then its low frequency is puzzling. It has been shown that the effective temperature of the WD is well correlated with the amplitude and period of the pulsations in isolated ZZ~Ceti WDs \citep{Clemens93,Mukadam06}. The lowest frequency pulsations are seen in the coolest WDs (e.g. G29-38, \citealt{Kleinman98}). In the case of GW~Lib therefore, we would expect that as the WD cools and enters the CV instability region we would see higher frequency pulsations develop, consistent with the high WD temperature. From the correlation shown in figure 1 of \citet{Mukadam06}, the $\sim$$75$ cycles/day implies a WD temperature of $<11,000$K. This is clearly not the case, since spectral fits of GW Lib have shown the WD to be hotter than this during quiescence ($14,700$K, \citealt{Szkody02}). If the $\sim$$75$ cycles/day pulsation is associated with the WD, then this suggests that the correlation between effective temperature and period does not apply to GW Lib (and potentially other CV pulsators), probably due to the chemical composition of the accreted outer layers. It is possible that this signal is a DBV pulsation driven by the ionisation of helium \citep{Arras06}. We see a second new signal develop in the June 2008, with a frequency of $\sim$$292$ cycles/day. This frequency does not correspond to any known modes, or any of the predictions of \citet{Townsley04}.

It is unclear as to when the $\sim$$75$ cycles/day periodicity first became apparent. We see in Figure \ref{fig:gwlib_ampspec_jun07} that there is some evidence for a signal at a similar frequency in the July $2007$ data, although given the amplitude of this peak we suspect it is just flickering in the source. The signal is very clear at the end of March $2008$ but there is evidence for its presence from when our $2008$ observations began. However, the signal is significantly weaker at this point so we suggest that our March observations are very close in time to the first manifestation of this periodicity in the source emission. In the case of the $\sim$$292$ cycles/day pulsation, further monitoring will be necessary in order to determine its persistence.

\section{CONCLUSIONS}
\label{sec:conc}

In this paper we report observations of two CV pulsators: GW Lib and SDSS 1610. We took multi-band, high time-resolution observations of both sources in quiescence in May 2005, using the high speed CCD photometer ULTRACAM mounted on the VLT. We supplemented this with additional data from the University of Cape Town photometer mounted on the 1.9m telescope at SAAO. In both sources we resolve the dominant periods which have been observed by previous authors. In SDSS 1610 we do detect some additional lower amplitude signals: the large collecting area of the VLT provides a distinct advantage over previous studies for SDSS 1610, which is much less luminous than GW Lib and is a more challenging target for high time-resolution photometry. The VLT does not provide the same advantage over previous studies in the case of GW Lib, since in this source the accretion-driven flickering of the source is the limiting factor in further mode identifications. We find in both sources that the signals tend to be stronger towards the blue end of the visible spectrum. Of particular significance is the finding that the two principal modes in SDSS 1610 have a different colour dependence. This may be evidence that these modes are spherical harmonics with different $l$ numbers. Further multi-band observations of this source could confirm this. We note also that our frequency determination of the signal identified as the $2 f_1$ combination by previous authors suggests that this identification may be incorrect and this period is an independent mode.  

We took additional observations of GW Lib in June 2007 with VLT+ULTRACAM, which we supplemented with data from the William Herschel Telescope taken a month later. These observations were made in the aftermath of an outburst in this source: the first outburst observed since its discovery and the first outburst observed in a CV known to contain a pulsating white dwarf. We believe that at the time of our observations the emission from the source is dominated by the white dwarf, and we find it to be more than a magnitude brighter than in our previous observations due to heating by the outburst. We observe much short-timescale variation in this source but we do not observe the coherent pulsations we detected in quiescence, leading us to believe that these have been suppressed by the heating of the white dwarf. Our results suggest the heating of the white dwarf has pushed it outside of the instability region in the $T_{eff}$ -– $\log g$ plane. We observed this source again eleven months after outburst with LT+RATCam. The WD is still significantly hotter than during quiescence. We still do not observe the known modes, but we report the emergence of two new periodicities. The first was apparent in March/April 2008 with a frequency of $74.86 \pm 0.68$ cycles/day ($P = 1154$s) and a $g'$-band amplitude of $2.20$\% $\pm 0.18$. We observe the second in June 2008, with a frequency of $292.05 \pm 1.11$ cycles/day ($P=296$s) and a $g'$ amplitude of $1.25$\% $\pm 0.18$. 

In GW Lib, we observe an additional modulation in luminosity with a period of $\sim$$2.1$h. This has been detected before, but not in all previous observations. The origin of this modulation is unclear: it is apparently unrelated to the orbital period. We find this modulation to vary over the course of our observations in phase and/or period. We suggest that this is an accretion related phenomenon associated with the accretion disc. A similar variation is apparent in some of the post-outburst data, but we believe this is most likely to be an orbital modulation.

\section*{ACKNOWLEDGEMENTS}
CMC and TRM are supported under grant ST/F002599/1 from the Science and
Technology Facilities Council (STFC). ULTRACAM and SPL are supported by STFC grants PP/D002370/1 and PP/E001777/1.  PAW's research is supported by the National Research Foundation of South Africa and by the University of Cape Town. BW's research is supported by the University of Cape Town. DS acknowledges the support of a STFC Advanced Fellowship. The results presented in this paper are based on observations made with ESO Telescopes at the Paranal Observatory under programme IDs 075.D-0311 and 279.D-5027, observations made with the William Herschel Telescope operated on the island of La Palma by the Isaac Newton Group in the Spanish Observatorio del Roque de los Muchachos of the Instituto de Astrofisica de Canarias, and observations made with the 1.9m telescope operated by the South African Astronomical Observatory. The Liverpool Telescope is operated on the island of La Palma by Liverpool John Moores University in the Spanish Observatorio del Roque de los Muchachos of the Instituto de Astrofisica de Canarias with financial support from the UK Science and Technology Facilities Council. We also acknowledge with thanks the variable star observations from the AAVSO International Database contributed by observers worldwide and used in this research. This research has made use of NASA's Astrophysics Data System Bibliographic Services and the SIMBAD data base, operated at CDS, Strasbourg, France. Thanks also to Lars Bildsten and Dean Townsley for interesting dicussions, and to the anonymous referee for detailed comments which have led to significant improvements to this paper.

\bibliography{gwlib_1610}

\end{document}